\documentclass[12pt]{article}
\usepackage[left=2.5cm,top=2.50cm,right=2.5cm,bottom=2.50cm]{geometry}
\usepackage{mathrsfs}
\usepackage{amsmath,amssymb,latexsym,color,cancel,graphicx,bbm,colortbl}
\usepackage[english]{babel}
\usepackage[latin1]{inputenc}
\usepackage{ragged2e}
\begin{document}
\date{}

\title{Non-Hermitian inverted Harmonic Oscillator-Type Hamiltonians Generated from Supersymmetry with Reflections}

\author{R. D. Mota$^{a}$, D. Ojeda-Guill\'en$^{b}$, M. Salazar-Ram{\'i}rez$^{b}$ \\ and V. D. Granados$^{c}$} \maketitle

\begin{minipage}{0.9\textwidth}

\small $^{a}$ Escuela Superior de Ingenier\'ia Mec\'anica y
El\'ectrica, Unidad Culhuac\'an, Instituto Polit\'ecnico Nacional,  Av. Santa Ana No. 1000, Col. San
Francisco Culhuac\'an, Del. Coyoac\'an,  C.P. 04430,  Ciudad de M\'exico, Mexico.\\

\small $^{b}$ Escuela Superior de C\'omputo, Instituto Polit\'ecnico Nacional, Av. Juan de Dios B\'atiz, Esq. Av. 
Miguel Oth\'on de Mendiz\'abal, Col. Lindavista, Del. Gustavo A. Madero, C.P. 07738,  Ciudad de M\'exico, Mexico.\\

\small $^{c}$ Escuela Superior de F\'{\i}sica y Matem\'aticas,
Instituto Polit\'ecnico Nacional,
Ed. 9, Unidad Profesional Adolfo L\'opez Mateos, Del. Gustavo  A. Madero, C.P. 07738, Ciudad de M\'exico, Mexico.\\
\end{minipage}

\abstract{By modifying and generalizing known supersymmetric models we are able to find four different sets of one-dimensional Hamiltonians for the inverted harmonic oscillator. The first  set of Hamiltonians is derived by extending the supersymmetric  quantum mechanics with reflections to non-Hermitian supercharges.  The second  set is obtained by generalizing the  supersymmetric  quantum mechanics valid for non-Hermitian supercharges with the Dunkl derivative instead of $\frac{d}{dx}$. Also, by changing the derivative $\frac{d}{dx}$ by the Dunkl derivative in the creation and annihilation-type operators of the standard inverted Harmonic oscillator 
$H_{SIO}=-\frac{1}{2}\frac{d^2}{dx^2}-\frac{1}{2}x^2$, we generate the third  set of Hamiltonians. The fourth  set  of Hamiltonians  emerges by allowing a parameter of the supersymmetric two-body Calogero-type model to take imaginary values. The eigensolutions of definite parity for each set of Hamiltonians are given.\\
\linebreak

\maketitle

\section{Introduction}

In 1951 Wigner \cite{wigner} introduced the deformed harmonic oscillator Heisenberg algebra in a non-explicit way. It was Yang  \cite{yang} who realized the Heisenberg algebra explicitly in terms of reflection operators.  Hamiltonians with reflection operators arose in the Calogero-Sutherland systems-type and their generalizations \cite{calogero, sutherland,genera}. The Hamiltonians of these systems are expressed in terms 
of the differential-difference exchange operators, known as Dunkl operators \cite{dunkl}. On the other hand, Witten \cite{witten} introduced the supersymmetric quantum mechanics  $SUSY$ $QM$ which has been found
many applications \cite{susy1,susy2}. Supersymmetric quantum mechanics with reflections  $SUSY$ $QM$-$R$ has been studied  
in the context of deformed Heisenberg algebras in Refs. \cite{plyuschay1, plyuschay2} and in a more systematical way in Ref.
\cite{post}. It  has been applied successfully to many physical systems as for example, the extended Scarf I potential \cite{post},  linear differential field equations \cite{plu1}, Aharonov-Bohm bound states, Dirac delta and P\"oschl-Teller potentials \cite{plu2}, bosons, fermions and anyons in the plane \cite{plu3}, anyons on the non-commutative plane \cite{plu4},  the two dimensional harmonic oscillator \cite{genest1}, and the Dunkl-Coulomb problem \cite{genest2}. Also, in Refs. \cite{correa1} and \cite{correa2} the hidden (bosonized) supersymmetry in finite-gap Lam\'e and Lam\'e associated systems was discovered and the involved physics was discussed. The superconformal symmetry of the quantum harmonic oscillator was explained in Refs. \cite{jaku} and \cite{inzunza}. In Ref. \cite{correa3} the bosonized supersymmetry of pairs of quantum systems with complexified Scarf II potentials was studied.

Another important branch to study quantum systems is the so-called non-hermitian quantum mechanics. The main problem is to find criteria for a non-Hermitian Hamiltonian to have a real spectrum. The requirement of quantum mechanics that the Hamiltonian $H$ be Hermitian in order to guaranties that the energy spectrum is real and that time evolution is unitary has to be changed. In non-Hermitian quantum mechanics it was found that the criteria for a quantum Hamiltonian to have a real spectrum is that it possesses an unbroken $\mathcal{P}\mathcal{T}$ symmetry ($\mathcal{P}$ is the space-reflection operator or parity operator, and $\mathcal{T}$ is the time-reversal operator) \cite{bender}.    An alternative approach to explore the basic structure responsible for the reality of the spectrum of a non-Hermitian Hamiltonian is by the notion of the pseudo-hermiticity introduced in Ref. \cite{mosta}. Both approaches  to non-Hermitian quantum mechanics were applied to nuclear physics, condensed
matter physics, relativistic quantum mechanics and quantum field theory, quantum cosmology, electromagnetic
wave propagation, open quantum systems, magnetohydrodynamics, quantum chaos, and biophysics \cite{bender,mosta}.

The exactly solvable Hamiltonian $H_{SIO}=-\frac{1}{2}\frac{d^2}{dx^2}-\frac{1}{2}x^2$ is known as  the inverted oscillator, repulsive oscillator, inverse oscillator or repulsive barrier. We will refer to it as the standard inverted oscillator ($SIO$) Hamiltonian.  It was introduced as an exercise in the Landau's classical book on quantum mechanics \cite{landau} and its physical applications began  after the Barton's publication of his Ph.D. thesis results \cite{barton}.  It is a very interesting model not only for being an exact quantum mechanical potential  but because of its recent applications in physics. It has been used among others,  as a model of instability in quantum mechanics \cite{shimbori1},  in the study of chaotic systems \cite{miller, gaioli}, and to study the statistical fluctuations of fusion dynamics \cite{hofmann}.  The inverted oscillator has been treated with group theory \cite{kalnins,wolf1},   by means of the Wigner function in the quantum phase space  \cite{wolf2},  by the Dirac-type operator method \cite{shimbori1}, by supersymmetric quantum mechanics \cite{shimbori2}, and more recently from the first and second-order factorization methods \cite{fernandez}. It is worth noticed the equivalence relation between free particle and harmonic oscillator. In fact, the time-dependent solutions of the free-particle Sch\"odinger equation can be mapped to those of the  Sch\"odinger equation for the harmonic potentials. In this regard  see Refs. \cite{takagi,ole} for the standard harmonic oscillator and  Ref. \cite{yuce} for the inverted harmonic oscillator.   It must be emphasized the prominent role of the inverted oscillator to study of the early time evolution in inflationary models \cite{guth}, in inflation and squeezed states \cite{andreas},
in describing the $D$-branes decay \cite{janik} and in string theory (see for example  Cap. 12, Sec. 8 of Ref.  \cite{zwiebach}).

In the present paper we introduce an interesting link between supersymmetric quantum mechanics with reflections, non-Hermitian Hamiltonians and  new realizations of the inverted harmonic oscillator systems. We generate from supersymmetry four sets of inverted harmonic oscillator-type Hamiltonians, which are generalizations of $H_{SIO}$ (Hermitian Hamiltonian) and its solution. It is well known that the $SIO$ is a genuine scattering Hamiltonian with real spectrum. We show that this genuine feature of  $H_{SIO}$ remains in the Hamiltonians generated from $SUSY$ $QM$-$R$ but their non-Hermiticity imposes a complex spectrum.

The paper is organized as follows.  In Section 2, we generalize the  $SUSY$ $QM$-$R$  Hermitian supercharges of Ref. \cite{post} to non-Hermitian ones. By setting properly the two functions involved in the model we generate the first set of $SUSY$ Hamiltonians for the inverted oscillator. 
In Section 3, the second set of Hamiltonians is obtained by generalizing the  $SUSY$ $QM$-$R$ to non-Hermitian supercharges with the Dunkl derivative $D=\frac{d}{dx}+\frac{\mu}{x}-\frac{\mu}{x}R$ instead of $\frac{d}{dx}$.
In Section 4, the factorization of the standard inverted Harmonic oscillator
$H_{SIO}$  is generalized by using the Dunkl derivative $D$ instead of $\frac{d}{dx}$. This procedure allows us to obtain the third set of the inverted harmonic oscillator Hamiltonians.  In Section 5,  we use the model of Ref. \cite{minimal}, but allowing one of its parameters to take the values $\pm i$ instead of $\pm 1$. This let us to obtain the fourth set  
of Hamiltonians for the inverted oscillator. The eigensolutions of definite parity for each set of Hamiltonians are given. Finally, we give the concluding remarks.               

\section{ The supersymmetric inverted harmonic oscillator with reflections}
 
In this Section we modify the formulation of the  $SUSY$ $QM$-$R$  given in Ref. \cite{post} applied to Hamiltonians which possesses only one supercharge $Q$, such that $H=Q^2$. To this end, we propose as supercharges the operators 
\begin{equation}
Q_\epsilon=\frac{1}{\sqrt{2}}\left(\frac{d}{dx}+U(x)\right)R+\frac{\epsilon}{\sqrt{2}}V(x),
\end{equation}  
being $U(x)$ and $V(x)$ even and odd functions, respectively, and  $R$ the Hermitian reflection operator $Rf(x)=f(-x)$. It is immediate to show that $R$ possesses the properties $R^2=1$,  $\frac{d}{dx}R=-R\frac{d}{dx}$ and $Rx=-xR$.

By direct calculation we obtain the supersymmetric Hamiltonian 
\begin{equation}
H_\epsilon=Q_\epsilon^2=-\frac{1}{2}\frac{d^2}{dx^2}+\frac{1}{2}\left(U^2+\epsilon^2V^2\right)+\frac{1}{2}\frac{dU}{dx}-\frac{\epsilon}{2}\frac{dV}{dx}R.
\end{equation}
Notice that if $\epsilon$ is real, then $Q_\epsilon$ and $H_\epsilon$ are Hermitian, whereas if $\epsilon$ is a complex number or a purely imaginary quantity, neither $Q_\epsilon$ nor $H_\epsilon$ are Hermitian. If $\epsilon=1$, the original model is recovered.  In the present paper we restrict $\epsilon$ to take the purely imaginary values, $\epsilon=i$ or $\epsilon=-i$. \\   
\linebreak 
Case I) $\epsilon= i$. 

If we set $V(x)=x$ and $U(x)=0$, then, the non-Hermitian supercharge is 
\begin{equation}
Q_1=\frac{1}{\sqrt{2}}\left(\frac{d}{dx}R+ix\right).
\end{equation}
Therefore, it generate the supersymmetric Hamiltonian 
\begin{equation}
H_1=Q_1^2=-\frac{1}{2}\frac{d^2}{dx^2}-\frac{1}{2}x^2-\frac{i}{2}R\label{ham}.
\end{equation}
If we introduce the creation and annihilation-type operators 
\begin{equation}
a_1^-=\frac{1}{\sqrt{2}} \left(\frac{d}{dx}+ix\right),\hspace{6ex}a_1^+=\frac{1}{\sqrt{2}} \left(-\frac{d}{dx}+ix\right),\label{creacion}
\end{equation} 
this Hamiltonian can be written as 
\begin{equation}
H_1=a_1^- a_1^+-\frac{i}{2}\left(1+R\right)=a_1^+ a_1^-+\frac{i}{2}\left(1-R\right).
\end{equation}
We point out that the operators (\ref{creacion}) are not adjoint each other, this is why we denoted them as $a_1^-$ and $a_1^+$, according to the notation used in Ref. \cite{fernandez}. \\ 
\linebreak
Since the eigenvalues of the reflection operator are $\pm 1$, then the Schr\"odinger equation eigensolutions  $H_1^\pm\psi_1^\pm (x)=E_1 ^\pm \psi_1^\pm(x)$, which are regular at the origin and possess definite parity $R\psi_1^\pm (x)=\pm \psi_1^\pm (x)$, are   
\begin{eqnarray}
\psi_1^\pm(x)=Ce^{-i\frac{x^2}{2}}x^{\frac{1}{2}\mp\frac{1}{2}}\hspace{1ex} {_1 F_1\left( \frac{1}{2}\mp \frac{1}{2}+i\frac{E_1^\pm}{2},1\mp \frac{1}{2};\,i{x}^{2}\right)}.\label{sol1}
\end{eqnarray}
It is well known that the  confluent hypergeometric function ${_1 F_1 (\lambda,\gamma,z)}$ converges if $|z|<\infty$, and the parameters $\lambda$ and $\gamma$ take arbitrary real or complex values, except $\gamma= 0, -1, -2,..., $. Thus, $\frac{1}{2}\mp \frac{1}{2}+i\frac{E_1^\pm}{2}=\lambda_1^\pm$, from which we obtain the definite parity energy spectrum for $H_1$:
\begin{equation}
E_1^\pm=i(1\mp 1-2\lambda_1^\pm),\hspace{5ex} \lambda_1^\pm\in \mathbb{C}.\label{e1}
\end{equation}

Case II) $\epsilon=- i$. 

With $V(x)=x$ and $U(x)=0$, we obtain the supercharge 
\begin{equation}
Q_2=\frac{1}{\sqrt{2}}\left(\frac{d}{dx}R-ix\right)
\end{equation}
and the supersymmetric Hamiltonian 
\begin{equation}
H_2=Q_2^2=-\frac{1}{2}\frac{d^2}{dx^2}-\frac{1}{2}x^2+\frac{i}{2}R.\label{h2}\\
\end{equation}
Due to the eigenvalues of the reflection operator, this Hamiltonian formally has  the same solutions as
those of the Hamiltonian (\ref{ham}). We point out  that the one dimensional standard harmonic oscillator with reflections  analogous to the 
inverted harmonic oscillator treated in this Section was studied in Refs. \cite{post} and \cite{minimal} in the context of  $N=\frac{1}{2}$ and $N=2$  $SUSY$ $QM$-$R$, respectively.

\section{The Dunkl inverted oscillator}

In this section we generalize the $SUSY$ $QM$-$R$ supercharges introduced in Section 2.  To this end, we change the standard derivative $\frac{d}{dx}$ by the Dunkl derivative  $D=\frac{d}{dx}+\frac{\mu}{x}-\frac{\mu}{x}R$, where as it is usual,  we restrict the parameter $\mu$ to take the values $\mu >-\frac{1}{2}$ \cite{genest1,minimal}. Thus,
\begin{equation}
Q_\epsilon=\frac{1}{\sqrt{2}}\left(D+U(x)\right)R+\frac{\epsilon}{\sqrt{2}}V(x),
\end{equation}  
and the Hamiltonian it generate is 
\begin{eqnarray}
&&H_\epsilon=Q_\epsilon^2=\frac{1}{2}\left\{-D^2+U^2+\epsilon^2V+[D,U]-\epsilon[D,V]R\right\}\nonumber \\
&&\hspace{4ex}=-\frac{1}{2}\frac{d^2}{dx^2}+\frac{1}{2}(U^2+\epsilon^2V^2)-\frac{\mu}{x}\frac{d}{dx}+\frac{\mu}{2x^2}(1-R)+\frac{1}{2}\frac{dU}{dx}-\frac{\epsilon}{2}\frac{dV}{dx}R.
\end{eqnarray}
If we set $\epsilon=i$, $V(x)=x$  and $U(x)=0$, we obtain the Hamiltonians 
\begin{equation}
H_{D1}=-\frac{1}{2}\frac{d^2}{dx^2}-\frac{1}{2}x^2-\frac{\mu}{x}\frac{d}{dx}+\frac{\mu}{2x^2}(1-R)- \frac{i}{2}R,\label{DD1}
\end{equation} 
whereas if we set  $\epsilon=-i$, $V(x)=x$  and $U(x)=0$, we get
 \begin{equation}
H_{D2}=-\frac{1}{2}\frac{d^2}{dx^2}-\frac{1}{2}x^2-\frac{\mu}{x}\frac{d}{dx}+\frac{\mu}{2x^2}(1-R)+ \frac{i}{2}R.\label{DD2}
\end{equation} 

Since the eigenvalues of $R$ are $\pm 1$, $H_{D1}$ and $H_{D2}$ possess essentially the same physical and mathematical properties. Thus for example, the definite parity eigensolutions,  finite at the origin for $H_{D1}$, are given by
\begin{equation}
\psi_{D1}^\pm (x)=Ce^{-i\frac{x^2}{2}} x^{\frac{1}{2}\mp \frac{1}{2}}\hspace{1ex} {_1 F_1\left(\frac{1}{2}\mp \frac{1}{2}+\frac{\mu}{2}+i\frac{E_{D1}^\pm}{2},1\mp \frac{1}{2}+\mu;\,i{x}^{2}\right)}. \label{sodd1}
\end{equation}
Using the same arguments we gave in Section 2, we find that the energy spectrum is given by 
\begin{equation}
E_{D1}^\pm=i(1\mp 1-\mu -2\lambda_{D1}^\pm),\hspace{5ex} \lambda_{D1}^\pm\in \mathbb{C}.\label{ed1}
\end{equation} 
It is important to note that if we set $\mu=0$ in the Hamiltonian (\ref{DD1}), it reduces to the Hamiltonian (\ref{ham}). This leads to the solutions (\ref{sodd1}) reduce to those of equation (\ref{sol1}), and the spectrum (\ref{ed1}) reduces properly to the spectrum (\ref{e1}), as it was expected.      

\section{Dunkl creation and annihilation-type operators and the inverted oscillator }

The standard inverted harmonic oscillator Hamiltonian can be written as  \cite{fernandez}
\begin{equation}
H_{SIO}=a_1^+ a_1^-+\frac{i}{2}=a_1^- a_1^+-\frac{i}{2}
\end{equation}  
with the operators $a_1^-$  and $a_1^+$ given by equation (\ref{creacion}).  However, as it will be shown below, when the derivative $\frac{d}{dx}$ is generalized to the Dunkl derivative $D$, the equality does not hold. 

Thus for $\epsilon=i$,  the Dunkl creation and annihilation-type operators for the inverted oscillator are
 \begin{eqnarray}
&&a_D^-=\frac{1}{\sqrt{2}} \left(D+ix\right)=\frac{1}{\sqrt{2}} \left(\frac{d}{dx}+\frac{\mu}{x}-\frac{\mu}{x}R+ix\right),\\
&&a_D^+=\frac{1}{\sqrt{2}} \left(-D+ix\right)=\frac{1}{\sqrt{2}}\left(-\frac{d}{dx}-\frac{\mu}{x}+\frac{\mu}{x}R+ix\right).
\end{eqnarray}
Also, as in the case of the operators of equation (\ref{creacion}), these operators are not adjoint each other.  
Hence, we obtain the Hamiltonians
\begin{eqnarray}
&&H_{CD1}\equiv a_D^+ a_D^-+\frac{i}{2} \nonumber \\
&&\hspace{3ex}=-\frac{1}{2}\frac{d^2}{dx^2}-\frac{1}{2}x^2 -\frac{\mu}{x}\frac{d}{dx}+\frac{\mu}{2x^2}\left(1-R\right)-i\mu R, \label{CD1}
\end{eqnarray}
and 
\begin{eqnarray}
&&H_{CD2}\equiv a_D^- a_D^+ -\frac{i}{2} \nonumber\\
&&\hspace{3ex}=-\frac{1}{2}\frac{d^2}{dx^2}-\frac{1}{2}x^2 -\frac{\mu}{x}\frac{d}{dx}+\frac{\mu}{2x^2}\left(1-R\right)+i\mu R,\label{CD2}
\end{eqnarray}
which result to be different because of the last term. However,  due to the  possible eigenvalues of the reflection operator, we consider these Hamiltonians as equal. Hence, for example,  we find that the eigenfunctions, finite at the origin and of definite parity $R\psi_{CD1}^\pm=\pm\psi_{CD1}^\pm$ of  $H_{CD1}^\pm$ (with eigenenergy $E_{CD1}^\pm$) are given by
\begin{equation}
\psi_{CD1}^\pm(x)=Ce^{-i\frac{x^2}{2}} x^{\frac{1}{2}\mp\frac{1}{2}}\hspace{1ex} {_1 F_1\left(\frac{1}{2}\mp\frac{1}{4}+\left(\frac{1}{2}\mp\frac{1}{2}\right)\mu+i\frac{E_{CD1}^\pm}{2},1\mp\frac{1}{2}+\mu;\,i{x}^{2}\right)}.
\end{equation}
In this case, we find that the energy spectrum is  
\begin{equation}
E_{CD1}^\pm=i\left(1\mp \frac{1}{2}+(1\mp 1)\mu -2\lambda_{CD1}^\pm\right),\hspace{5ex} \lambda_{CD1}^\pm\in \mathbb{C}.\label{ecd1}
\end{equation}  
Notice that our results above in this Section were found for $\epsilon=1$.  Completely analogous results are derived for $\epsilon=-i$. Also, we remark that although the Hamiltonian $H_{D1}$  of Section 3 and $H_{CD1}$ are very similar, their corresponding eigensolutions are not.  We emphasize that our Hamiltonians and their solutions of the present Section are  properly reduced for $\mu=0$ to those solutions of definite parity for the simplest  inverted oscillator (for the $H_{SIO}$ Hamiltonian) \cite{fernandez}.   

\section{Supersymmetric two-body Calogero -type model}

We use the supersymmetric version of the two-body Calogero model of Ref. \cite{minimal},
\begin{eqnarray}
H_c=-\frac{1}{2}\left(\frac{d}{dx}+\left(\epsilon x-\frac{\mu}{x}\right)R\right)^2\hspace{19ex}\nonumber\\
\hspace{10ex}=-\frac{1}{2}\frac{d^2}{dx^2}+\frac{\epsilon^2}{2}x^2+\frac{1}{2}\frac{\mu^2}{x^2}-\frac{\mu}{2x^2}R-\frac{\epsilon}{2}(2\mu+R),\label{calog}
\end{eqnarray}
to the inverted oscillator. The original model was applied to study the standard one-dimensional harmonic oscillator by setting  $\epsilon=\pm 1$. In order to obtain Hamiltonians of the inverted oscillator-type, we set $\epsilon = \pm i$. To be specific, as it is usual,  we will assume that the parameter $\mu$ satisfy $\mu>-\frac{1}{2}$ \cite{genest1,minimal}.   Also, we take the two possibilities for the reflection operator eigenvalues,  $\pm 1$.  Thus, for $\epsilon= i$,  we obtain the Calogero-type supersymmetric Hamiltonians
\begin{equation}
H_{c1}^\pm=-\frac{1}{2}\frac{d^2}{dx^2}-\frac{1}{2}x^2 +\frac{\mu^2}{2x^2}\mp \frac{\mu}{2x^2}-\frac{i}{2}(2\mu \pm 1).\label{hc1}\\
\end{equation} 
We find that the regular eigenstates at the origin of well-defined parity of this Hamiltonian are given by
\begin{equation}
\psi_{c1}^\pm (x)=C_1e^{-i\frac{x^2}{2}} |x|^\mu x^{\frac{1}{2}\mp \frac{1}{2}}\hspace{1ex} {_1 F_1\left(\frac{1}{2}\mp \frac{1}{2}+i\frac{E_{c1}^\pm}{2},1\mp \frac{1}{2}+\mu;\,i{x}^{2}\right)}\label{s1},
\end{equation}
with energy spectrum 
\begin{equation}
E_{c1}^\pm=i(1\mp 1-2\lambda_{c1}^\pm), \hspace{5ex} \lambda_{c1}^\pm\in \mathbb{C}.\label{ec1}
\end{equation}
With $\epsilon=-i$, we obtain the Hamiltonians 
\begin{equation}
H_{c2}^\pm=-\frac{1}{2}\frac{d^2}{dx^2}-\frac{1}{2}x^2 +\frac{\mu^2}{2x^2}\mp \frac{\mu}{2x^2}+\frac{i}{2}(2\mu \pm1). \label{hc2}
\end{equation} 
The even end odd eigenstates and their corresponding energies are 
\begin{equation}
\psi_{c2}^\pm (x)=C_{2}e^{-i\frac{x^2}{2}}  |x|^\mu x^{\frac{1}{2}\mp \frac{1}{2}}\hspace{1ex} {_1 F_1\left(\frac{1}{2}+ \mu+i\frac{E_{c2}^\pm}{2},1\mp\frac{1}{2}+\mu;\,i{x}^{2}\right)}\label{s2},
\end{equation} 
and 
\begin{equation}
E_{c2}^\pm=i(1+2\mu-2\lambda_{c2}^\pm),\hspace{5ex} \lambda_{c2}^\pm\in \mathbb{C}.\label{ec2}
\end{equation}
We observe that the Hamiltonian (\ref{calog}) with $\epsilon=i$ and $\mu=0$, reduces to the Hamiltonian (\ref{ham}) as a particular case.  Consistently, for the same conditions, the eigenfunctions (\ref{s1}) reduce to those of equation (\ref{sol1}) and the spectrum (\ref{ec1}) reduces to the spectrum (\ref{e1}).
Since the Hamiltonian (\ref{calog})  with $\epsilon=-i$ and  $\mu=0$ reduces  to the Hamiltonian (\ref{h2}), this reduction is also satisfied between their definite parity solutions and their spectrum.

We point out that the Hamiltonians (\ref{ham}),  (\ref{h2}),  (\ref{DD1}),  (\ref{DD2}), (\ref{CD1}), (\ref{CD2}), (\ref{hc1}) and (\ref{hc2}) generated from $SUSY$ $QM$-$R$ are non-Hermitian because the imaginary unit accompanies the reflection operator.  
 As in the case of the simplest inverted oscillator, whose eigenfunctions are not $\mathcal{L}^2(\mathbb{R})$-square integrable for an arbitrary energy  $E\in \mathbb{C}$  \cite{shimbori1,yuce, fernandez},
the Hamiltonians and their definite parity eigenfunctions reported  in this paper result to be non-$\mathcal{L}^2(\mathbb{R})$-square integrable.  As a consequence of the properties of the confluent hypergeometric function, equations (\ref{e1}),  (\ref{ed1}), (\ref{ecd1}), (\ref{ec1}) and (\ref{ec2})  show that the spectrum of our different Hamiltonians is complex.

\section{Concluding Remarks}

In this paper we have used and modified different approaches of supersymmetry with reflections to generate four sets of  one-dimensional Hamiltonians for the inverted harmonic oscillator. In our opinion, the algebraic properties of the inverted harmonic oscillator manifested in this work are inherited from those of the standard harmonic oscillator. This is in agreement  with the mathematical similarity between both systems, although they are physically very different.

Relating to the solutions we have reported in this paper,  it is necessary the following remark. We focus our attention to the standard inverted oscillator Hamiltonian  $H_{SIO}$, which is Hermitian and possesses, in general,  eigensolutions which are non-$\mathcal{L}^2(\mathbb{R})$-square integrable for an arbitrary energy $E\in \mathbb{C}$.  
In Refs. \cite{wolf1,fernandez} it has been shown that the correct spectrum of $H_{SIO}$ is $E\in\mathbb{R}$ and its correct eigenfunctions are a suitable linear combination of the even and odd eigenfunctions.
This problem does not exist for the definite parity solutions of the non-Hermitian Hamiltonians of the present paper. We showed that the properties of the confluent hypergeometric function allow the energy spectrum to be complex, and the scattering nature of our Hamiltonians   is reflected in the non-$\mathcal{L}^2(\mathbb{R})$-square integrable eigenfunctions for an arbitrary energy $E\in \mathbb{C}$.

We emphasize that our results have two main contributions. First one, in extending $SUSY$ $QM$-$R$ models to the case of non-Hermitian Hamiltonians. Second one:  their application to generate different sets of non-Hermitian inverted harmonic oscillator Hamiltonians. We consider that our results are an important contribution to the study of non-Hermitian Hamiltonians and to the study of the inverted harmonic oscillator, which possesses a long-range of applicability, as it was emphasized in the introduction. 
        
Finally, we mention that in Ref. \cite{lohe} the one-dimensional harmonic oscillator in fractional dimensions was studied. This  problem could be treated  in a direct  way with the methods introduced in the present paper.

\section*{Acknowledgments}
This work was partially supported by SNI-M\'exico, COFAA-IPN, EDI-IPN, EDD-IPN and
CGPI-IPN Project Numbers 20180741,  20181711  and 20180245.

\end{document}